\documentclass[conference]{IEEEtran}
\IEEEoverridecommandlockouts

\usepackage{graphicx}
\usepackage{amsmath}
\usepackage{amsfonts}
\usepackage{amssymb}
\usepackage{amsthm}
\usepackage{braket}
\usepackage{listings}
\usepackage{subcaption}
\usepackage{xcolor}
\usepackage{wrapfig}
\usepackage{url}
\usepackage{inconsolata} 
\usepackage[utf8]{inputenc}
\usepackage[switch]{lineno}
\usepackage{microtype}

\newcommand*\circled[1]{\ifcase#1\or ①\or ②\or ③\or ④\or ⑤\or ⑥\or ⑦\or ⑧\or ⑨\fi}

\definecolor{codegray}{gray}{0.95}
\definecolor{codeblue}{rgb}{0.0, 0.0, 0.6}
\definecolor{codegreen}{rgb}{0.0, 0.5, 0.0}
\definecolor{codepurple}{rgb}{0.58, 0.0, 0.82}
\definecolor{codered}{rgb}{0.6, 0.0, 0.0}

\lstdefinelanguage{Rust}{
  keywords={typeof, new, true, false, catch, function, return, null, catch, switch, var, if, in, while, do, else, case, break, const, let, mut, match, fn, pub, impl, struct, enum, use, mod, crate, where, unsafe, async, await, move, type, trait, self, Self},
  keywordstyle=\color{blue}\bfseries,
  ndkeywords={class, export, boolean, throw, implements, import, this},
  ndkeywordstyle=\color{darkgray}\bfseries,
  identifierstyle=\color{black},
  sensitive=true,
  comment=[l]{//},
  morecomment=[s]{/*}{*/},
  commentstyle=\color{gray}\ttfamily,
  stringstyle=\color{black}\ttfamily,
  morestring=[b]',
  morestring=[b]"
}

\lstdefinestyle{livelypython}{
    language=Python,
    backgroundcolor=\color{codegray},
    basicstyle=\footnotesize\ttfamily,
    keywordstyle=\color{blue}\bfseries,
    stringstyle=\color{codepurple},
    commentstyle=\color{codegreen}\itshape,
    identifierstyle=\color{black},
    numberstyle=\tiny\color{darkgray},
    morekeywords={QubitGate, DifferentiableUnitary, UnitaryMatrix, np},
    frame=single,
    rulecolor=\color{black!30},
    frameround=tttt,
    breaklines=true,
    showstringspaces=false,
    escapechar=`
}

\lstdefinestyle{livelyrust}{
    language=Rust,
    backgroundcolor=\color{codegray},       
    basicstyle=\footnotesize\ttfamily,      
    keywordstyle=\color{codeblue}\bfseries,    
    stringstyle=\color{black},         
    commentstyle=\color{codegreen}\itshape,    
    identifierstyle=\color{black},            
    numberstyle=\tiny\color{darkgray},        
    morekeywords={usize, QuditCircuit, UnitaryExpression, f64, vec, fn, let, mut, for, in},
    frame=single,                           
    rulecolor=\color{black!30},              
    frameround=tttt,                        
    breaklines=true,
    literate={θ}{{$\theta$}}1 {ϕ}{{$\phi$}}1 {λ}{{$\lambda$}}1,
    showstringspaces=false,
}

\theoremstyle{definition}

\newcommand{\meta}[1]{\textit{#1}}
\newcommand{\term}[1]{\texttt{#1}}
\def\BibTeX{{\rm B\kern-.05em{\sc i\kern-.025em b}\kern-.08em
    T\kern-.1667em\lower.7ex\hbox{E}\kern-.125emX}}

\title{OpenQudit: Extensible and Accelerated Numerical Quantum Compilation via a JIT-Compiled DSL}

\author{
    \IEEEauthorblockN{Ed Younis}
    \IEEEauthorblockA{
        \textit{Computational Research Division} \\
        \textit{Lawrence Berkeley National Laboratory} \\
        Berkeley, USA \\
        edyounis@lbl.gov
    }
}

\begin{document}
\maketitle

\begin{abstract}
High-performance numerical quantum compilers rely on classical optimization, but are limited by slow numerical evaluations and a design that makes extending them with new instructions a difficult, error-prone task for domain experts. This paper introduces OpenQudit, a compilation framework that solves these problems by allowing users to define quantum operations symbolically in the Qudit Gate Language (QGL), a mathematically natural DSL. OpenQudit's ahead-of-time compiler uses a tensor network representation and an e-graph-based pass for symbolic simplification before a runtime tensor network virtual machine (TNVM) JIT-compiles the expressions into high-performance native code. The evaluation shows that this symbolic approach is highly effective, accelerating the core instantiation task by up to $\mathtt{\sim}20\times$ on common quantum circuit synthesis problems compared to state-of-the-art tools.
\end{abstract}

\begin{IEEEkeywords}
Quantum computing, Compilers, Just-in-time (JIT) compilation, Symbolic computation, Tensor networks, Optimization
\end{IEEEkeywords}

\section{Introduction}

Quantum compilation is a critical step for executing programs on the diverse and highly constrained landscape of today's quantum processing units (QPUs). While many compilation strategies exist, numerical optimization-based methods~\cite{qsearch_qce20, bqskit, nacl, squander1, pam, xu2025optimizing} have emerged as a cornerstone technique, enabling the automatic discovery of high-quality, portable rewrite rules akin to classical superoptimization~\cite{superoptimization}. However, this reliance on numerical methods introduces two significant challenges. First, it creates a severe performance bottleneck, as the compiler must repeatedly evaluate complex mathematical functions and their gradients. Second, these compilers lack extensibility in practice; domain experts who wish to add support for emerging quantum operations must manually and efficiently implement often-complex analytical gradients, a process that is both difficult and error-prone.

To address these performance and extensibility challenges, this work introduces OpenQudit, a framework centered on the Qudit Gate Language (QGL). QGL is a domain-specific language designed to be mathematically natural, allowing experts to define quantum gate instructions as symbolic expressions with a syntax that mirrors on-paper formulations. The OpenQudit library then provides an extensible interface to compose these expressions into complex quantum programs and a high-performance pipeline to evaluate them. This pipeline leverages e-graphs~\cite{egraph} for symbolic simplification, LLVM~\cite{llvm} to perform just-in-time (JIT) compilation of these expressions into fast, native machine code for numerical evaluation, and a novel tensor network virtual machine to synthesize dedicated programs for efficient, repeated circuit evaluation. The resulting framework demonstrates performance speedups on average of $19.6\times$ for common 3-qubit quantum circuit synthesis problems compared to state-of-the-art tools.

This paper makes the following contributions:

\begin{itemize}
    \item \textbf{The Qudit Gate Language (QGL):} A mathematically natural DSL for describing qubit- and qudit-based quantum instructions.
    \item \textbf{A Novel Compilation Pipeline:} A system that uses e-graphs for symbolic simplification, LLVM for JIT compilation, and a tensor network virtual machine to synthesize highly optimized evaluation programs.
    \item \textbf{The OpenQudit Library:} An extensible and feature-rich interface for building complex quantum programs from QGL expressions.
    \item \textbf{A Comprehensive Evaluation:} A demonstration of optimization speedups over state-of-the-art numerical quantum compilation tools.
\end{itemize}

This paper is organized as follows. Section~\ref{sec:background} provides background on numerical quantum compilation. Section~\ref{sec:qgl} formally defines the Qudit Gate Language (QGL), and Section~\ref{sec:openqudit} details the OpenQudit system architecture. Section~\ref{sec:eval} presents the experimental evaluation, followed by a discussion of the results in Section~\ref{sec:disc}. The paper surveys related work in Section~\ref{sec:related} and concludes in Section~\ref{sec:conc}.

\section{Background and Motivation}
\label{sec:background}

\subsection{Quantum States and Operations}
The state of a $d$-level quantum system, or \textit{qudit}, is a superposition of $d$ basis states, denoted $\{\ket{0}, \ket{1}, \dots, \ket{d-1}\}$~\cite{nielsen2010quantum}. A pure qudit state $\ket{\phi}$ is a linear combination of these basis states, $\ket{\phi} = \sum_{i=0}^{d-1}{\alpha_i\ket{i}}$, where the complex coefficients $\alpha_i$, known as amplitudes, are constrained by $\sum|\alpha_i|^2 = 1$. The most common case is the \textit{qubit}, where $d=2$. A multi-qudit system is described by the tensor product of individual states, creating a larger composite state. While some of these composite states are separable, many are \textit{entangled}, meaning the state of the system cannot be described independently of its parts---a key resource in quantum computation.

Quantum programs transform these states using linear operators that preserve the pure state constraint. These operators are members of the unitary group $U(D)$, where $D$ is the total dimension of the system. While any unitary matrix is a valid transformation in theory, quantum hardware is constrained in practice to a small, fixed set of natively executable instructions or \textit{gates}. The term, gate, is often used interchangeably with quantum instruction, operation, or unitary. The U3 gate, a parameterized single-qubit gate, is a common instruction since it can parameterize any single-qubit unitary:

$$
    U3(\theta, \phi, \lambda) = \begin{bmatrix}
        \cos{\frac{\theta}{2}} & -e^{i\lambda}\sin{\frac{\theta}{2}} \\
        e^{i\phi}\sin{\frac{\theta}{2}} & e^{i(\phi + \lambda)}\cos{\frac{\theta}{2}}
    \end{bmatrix}
$$

\begin{figure}
    \centering
    \includegraphics[trim={1.8cm, 0.3cm, 1cm, 0.1cm},clip,width=\linewidth]{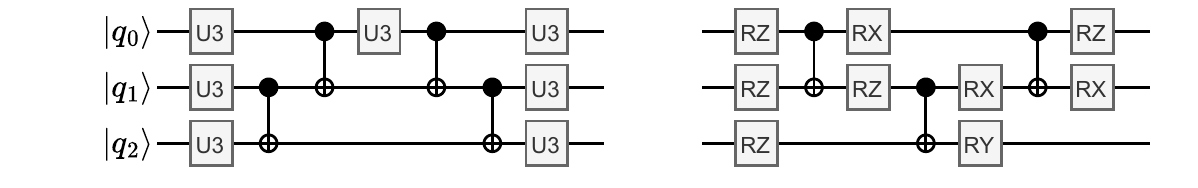}
    \caption{Two quantum programs illustrated in the circuit model. The left circuit comprises U3 and CNOT gates acting on three qubits. The starting state of the qubits is labeled on the left as $\ket{q_i}$; however, on the right, they are not labeled and are assumed to be indexed starting from the top and counting down. The right circuit uses a more diverse gate set that includes RX, RY, and RZ rotations.}
    \label{fig:circuit}
\end{figure}

Quantum \textit{programs} are typically expressed using the \textit{circuit model}~\cite{feynman1986quantum}, a diagramatic representation where wires horizontally extending from left to right represent the evolution of qudits through time. Gates, representing quantum operations, are placed on these wires. An entire circuit can be represented by a single unitary operator, $U_{\text{circuit}}$, calculated by taking the tensor product of all the individual gate operators in sequence. Figure~\ref{fig:circuit} provides an example of the model. A key task of a quantum compiler is \textit{circuit synthesis}: translating a high-level program description, often a large unitary matrix, into a sequence of native gates that a specific QPU can execute.

\subsection{Numerical Quantum Compilation}
A dominant paradigm in modern quantum compilation is \textit{numerical instantiation}~\cite{instantiation_qce22}, where a \textit{Parameterized Quantum Circuit (PQC)} is used to approximate a target unitary. A PQC is a circuit whose gates are parameterized by a set of real variables, $\vec\theta$. The compiler treats the PQC as a tunable function, $U(\vec\theta)$, and uses a classical, gradient-based optimizer to find the parameters $\vec\theta$ that best match a target operation, $U_{\text{target}}$. This is framed as minimizing a cost function, such as the infidelity:
\begin{equation}
L(\theta) = 1 - \frac{1}{D} |\text{Tr}(U_{\text{target}}^\dagger U(\vec\theta))|
\label{eq:infidelity}
\end{equation}
This technique is powerful and ubiquitous. For example, the state-of-the-art Berkeley Quantum Synthesis Toolkit (BQSKit)~\cite{bqskit} uses numerical instantiation to resynthesize sub-circuits, identifying shorter, more efficient gate sequences.

\subsection{The Performance and Extensibility Bottleneck}
While powerful, the reliance on numerical instantiation introduces two critical bottlenecks for compilers. First, the iterative nature of gradient descent creates a \textbf{performance challenge}, requiring repeated and computationally expensive evaluations of the cost function and its gradient.

Second, this approach suffers from poor \textbf{extensibility}. To enable gradient-based optimization, a compiler must provide the analytical gradient, $\frac{\partial U(\theta_i)}{\partial \theta_i}$, for each parameterized operation. For a domain expert who wishes to add a new gate, this requires deriving and manually implementing complex, error-prone matrix calculus. Listing~\ref{lst:bqskit_example} illustrates this challenge, showing the boilerplate code and separate, manually-derived gradient function required to add a single standard gate. This process creates a high barrier to entry. While hardware-compatible techniques like the parameter-shift rule~\cite{mitarai2018quantum} can compute gradients without analytical formulas, they often incur a substantial overhead, further slowing down the compilation process. This problem is exacerbated when moving to \textit{qudits}, where the higher dimensionality makes the analytical gradients significantly more complex and a standard set of gates has not yet been established.

\begin{figure}[t]
\begin{lstlisting}[
    style=livelypython,
    caption={
        An example of defining a new parameterized gate in a typical numerical compiler framework. To add the `U3` gate, a user must first, implement the required class boilerplate, provide a function to compute the gate's unitary matrix, and critically, manually derive and implement the complex analytical gradient. This process is labor-intensive and error-prone, creating a high barrier to extending the compiler.
    },
    label={lst:bqskit_example}
]
class U3Gate(QubitGate, DifferentiableUnitary):
  _num_qudits = 1
  _num_params = 3
  _qasm_name = 'u3'

  def get_unitary(self, params = []):
    ct = np.cos(params[0] / 2)
    st = np.sin(params[0] / 2)
    cp = np.cos(params[1])
    sp = np.sin(params[1])
    cl = np.cos(params[2])
    sl = np.sin(params[2])
    el = cl + 1j * sl
    ep = cp + 1j * sp

    return UnitaryMatrix(
      [
        [ct, -el * st],
        [ep * st, ep * el * ct],
      ],
    )

  def get_grad(self, params = []):
    ct = np.cos(params[0] / 2)
    st = np.sin(params[0] / 2)
    cp = np.cos(params[1])
    sp = np.sin(params[1])
    cl = np.cos(params[2])
    sl = np.sin(params[2])
    el = cl + 1j * sl
    ep = cp + 1j * sp
    del_ = -sl + 1j * cl
    dep_ = -sp + 1j * cp

    return np.array(
      [
        [
          [-0.5 * st, -0.5 * ct * el],
          [0.5 * ct * ep, -0.5 * st * el * ep],
        ],
        [
          [0, 0],
          [st * dep_, ct * el * dep_],
        ],
        [
          [0, -st * del_],
          [0, ct * ep * del_],
        ],
      ], dtype=np.complex128,
    )

\end{lstlisting}
\end{figure}

\begin{figure}[t]
\begin{lstlisting}[
    style=livelyrust, % Reusing the Rust style we defined
    caption={
        The same U3 gate from Listing~\ref{lst:bqskit_example} defined in a single, mathematically natural QGL expression. From this lone definition, OpenQudit automatically derives the unitary matrix, its analytical gradient, and the high-performance JIT-compiled code for both when needed.
    },
    label={lst:qgl_u3_example}
]
let u3 = UnitaryExpression::new("U3(θ, ϕ, λ) {
  [
    [cos(θ/2), ~e^(i*λ)*sin(θ/2)],
    [e^(i*ϕ)*sin(θ/2), e^(i*(ϕ + λ))*cos(θ/2)],
  ]
}");
\end{lstlisting}
\end{figure}

\begin{figure*}[t]
\centering
\begin{tabular}{lrcl}
  (1) & \meta{definition} & ::= & \term{ident} [ \meta{radices} ] \term{(} [ \meta{varlist} ] \term{)} \term{\{} \meta{expression} \term{\}} \term{;} \\
  (2) & \meta{radices} & ::= & \term{<} \meta{intlist} \term{>} \term{;} \\
  (3) & \meta{expression} & ::= & \meta{term} \; \{ (\term{+} $|$ \term{-}) \; \meta{term} \} \\
  (4) & \meta{term} & ::= & \{ \term{\textasciitilde} \} \; \meta{factor} \; \{ (\term{*} $|$ \term{/}) \; \meta{factor} \} \\
  (5) & \meta{factor} & ::= & \meta{primary} \; \{ \term{\textasciicircum} \; \meta{primary} \} \\
  (6) & \meta{primary} & ::= & \meta{variable} \;$|$\; \meta{constant} \;$|$\; \meta{function} \;$|$\; \meta{matrix} \;$|$\; \term{(} \meta{expression} \term{)} \\
  (7) & \meta{matrix} & ::= & \term{[} \meta{row} \{ \term{,} \meta{row} \} [ \term{,} ] \term{]} \\
  (8) & \meta{row} & ::= & \term{[} \meta{exprlist} \term{]} \\
\end{tabular}
\caption{The abstract syntax of the Qudit Gate Language (QGL). Metavariables are shown in italics, and terminals are in a monospaced font. Basic lexical and list productions are elided for clarity.}
\label{fig:qgl_grammar}
\end{figure*}

\subsection{Enabling Technologies in OpenQudit}
This work leverages two key technologies to overcome these bottlenecks. First, it uses \textit{e-graphs}, a data structure for efficiently representing equivalent expressions via equality saturation, to symbolically simplify QGL expressions and their gradients before compilation. 

Second, it utilizes \textit{tensor networks}, a graphical framework for representing complex chains of multi-linear algebra operations. A simple way to understand tensor networks is to view standard matrix multiplication as a specific case. The multiplication of two matrices, $C = AB$, is defined element-wise as $C_{ik} = \sum_j A_{ij} B_{jk}$. In the tensor network formalism, this corresponds to the contraction of two rank-2 tensors (matrices $A$ and $B$) over their shared index, $j$. The resulting tensor, $C$, has the remaining uncontracted, or ``open,'' indices, $i$ and $k$. Tensor networks generalize this fundamental operation, providing a powerful graphical notation for complex operations involving tensors of any rank and any pattern of contractions.

In the quantum context, tensor networks provide a framework for efficiently representing high-dimensional quantum objects like states and operators. A tensor network can also represent the contraction of all gates in a circuit, providing the most efficient method to evaluate a circuit's unitary. This representation is the foundation of the framework's tensor network virtual machine.


\section{QGL: The Qudit Gate Language}
\label{sec:qgl}

\subsection{Syntax and Semantics of QGL}

The \textit{Qudit Gate Language (QGL)} is a domain-specific language designed to represent quantum gate instructions as symbolic, unitary-valued expressions. Its design philosophy is to provide a mathematically natural syntax, allowing quantum experts to define complex operations in a manner that closely mirrors on-paper matrix formulations, without requiring deep systems programming expertise. In stark contrast to the verbose implementation required by traditional frameworks (Listing~\ref{lst:bqskit_example}), the same U3 gate is defined in QGL with a single, natural expression, as shown in Listing~\ref{lst:qgl_u3_example}.

The abstract syntax of QGL is formally defined in Figure~\ref{fig:qgl_grammar}. The grammar focuses on the core structural rules that define how gates are composed, while eliding standard lexical productions (e.g., for identifiers and constants) for clarity. At the top level, a \textbf{\meta{definition}} (1) binds a name to a symbolic expression. It may optionally specify the \textbf{\meta{radices}} (2) of the qudits it acts upon and a list of symbolic parameters. QGL expressions follow standard mathematical operator precedence, built hierarchically from \textbf{\meta{terms}} (3-4), \textbf{\meta{factors}} (5), and \textbf{\meta{primary}} expressions (6), which include variables, functions, and explicit matrix definitions (7-8).

QGL expressions are built from standard mathematical components. The language reserves the variables $i$, $e$, and $\pi$ for their usual mathematical values and provides a rich set of built-in functions, including common trigonometric (\texttt{sin}, \texttt{cos}, etc.), exponential (\texttt{exp}), and logarithmic (\texttt{ln}) functions. A key semantic constraint is that any expression must be symbolically expressible in a closed, element-wise form. This allows for straightforward support of matrix multiplication and addition, but excludes operations like the matrix exponential. A gate \meta{definition} can include an optional list of \meta{radices} (e.g., $<2,3>$ for a qubit-qutrit gate). If specified, the compiler verifies that the dimension of the resulting expression matrix matches the product of the radices. If omitted, the gate is assumed to operate only on qubits, and its dimension must be a power of two.

\subsection{Internal Representation and Composability}

After parsing, QGL definitions are lowered into an internal representation (IR) consisting of a 2D array of complex symbolic expressions. Each element in the array is a data structure containing separate symbolic trees for its real and imaginary parts, with all trigonometric functions canonicalized to \texttt{sin} and \texttt{cos} for uniform processing.

This symbolic IR is not static; it serves as a powerful foundation for programmatically constructing more complex operations. The OpenQudit library provides a suite of symbolic transformations that operate on this IR, including matrix multiplication, Kronecker product, substitution, and conjugation. This enables the flexible, on-the-fly creation of composite gates---such as controlled, inverted, or fused operations---directly from the user's high-level QGL definitions, forming the core of the framework's extensibility. Crucially, this symbolic IR is also fully differentiable. OpenQudit includes a symbolic differentiation engine that automatically computes the analytical gradient of any QGL expression. This approach replaces the manual, error-prone process of implementing gradients by hand. The result is another symbolic expression—representing the gradient—which can then be optimized alongside the original expression before code generation.

\subsection{Expression Optimization with E-Graphs}

The fast numerical optimization goal of OpenQudit requires fast evaluation of both a function and its gradient. As established, the symbolic IR provides the analytical gradient automatically, but this expression is often large and, in its initial construction, contains significant redundancies. Simplifying this symbolic gradient is therefore critical for performance. To achieve this, OpenQudit employs an e-graph-based simplification pass on both the original expression and its derivative before they are passed to the JIT compiler. E-graphs~\cite{egraph} and the technique of equality saturation are well-suited for this task, as they allow for the efficient exploration and representation of a vast space of semantically equivalent expressions. The implementation leverages the EGG library~\cite{egg}.

The effectiveness of equality saturation depends on a high-quality set of rewrite rules. The rule set in OpenQudit was bootstrapped from a foundational set of real-valued expression rules extracted from Herbie~\cite{herbie} and then refined and expanded using Enumo~\cite{enumo}. This process yielded a robust set of rules sufficient to discover all closed-form trigonometric identities available on Wikipedia.

The simplification process follows three main steps. First, an e-graph is populated with the symbolic expressions from both the real and imaginary components of a gate's unitary and its gradient. Second, equality saturation is run on the e-graph to find and represent all equivalent forms of these expressions based on the rewrite rule set. Since QGL expressions for individual gates are typically small and sparse, the e-graphs are not expected to become large. Nonetheless, standard safeguards are applied, including iteration and node-count limits, to prevent potential saturation blow-up. Third, a final, optimized expression is extracted from each e-class (a set of equivalent expressions) according to a custom cost function, detailed in Table~\ref{tab:eggcost}. This cost function is designed to encourage the use of trigonometric identities. As quantum gates are typically composed of numerous trigonometric functions, the primary objective is to reduce the final count of expensive $\sin$ and $\cos$ operations (without introducing other costly functions like $\ln$ or $\exp$) and to promote common subexpression elimination. The results are robust to small changes in the cost function, as the large separation between cheap arithmetic and expensive trigonometric operations is the dominant factor.

\begin{table}[h!]
    \centering
    \begin{tabular}{|l|c|}
        \hline
        \textbf{Expression Type} & \textbf{Cost} \\
        \hline
        $\pi$, Variable & 0.0 \\
        Constant & 0.5 \\
        $\mathtt{\sim}, \texttt{+}, \texttt{-}$ & 1.0 \\
        $\texttt{*}, \texttt{/}$ & 5.0 \\
        $\text{sqrt}, \sin, \cos$ & 50.0 \\
        $\exp, \ln, \text{pow}$ & 100.0 \\
        \hline
    \end{tabular}
    \caption{The cost function used for extracting simplified expressions from the e-graph.}
    \label{tab:eggcost}
\end{table}


While optimal expression extraction from an e-graph can be formulated as an Integer Linear Programming (ILP) problem~\cite{ilpextraction}, this approach is often too slow for a production compiler. Instead, OpenQudit implements a novel, greedy bottom-up extraction heuristic that offers a favorable trade-off between compilation time and expression quality. The algorithm first stabilizes costs across the e-graph by iteratively calculating the minimum cost for each e-class based on the current costs of its children. After stabilization, the lowest-cost expression is extracted. To explicitly encourage \textit{Common Subexpression Elimination (CSE)}, the cost of each e-class traversed during an extraction is immediately set to zero. This incentivizes subsequent extractions to reuse already-computed subexpressions. This cost-update and extraction process is repeated until all required expressions have been extracted from the e-graph. A clear example is the U2 gate, which contains the subexpressions $e^{i\lambda}$, $e^{i\phi}$, and $e^{i(\phi + \lambda)}$. The rewrite rules find the equivalent form $e^{i\lambda} \cdot e^{i\phi}$ for the last element. During extraction, once $e^{i\lambda}$ and $e^{i\phi}$ are extracted (and their costs set to zero), the heuristic will greedily select the $e^{i\lambda} \cdot e^{i\phi}$ form for the final element, as it efficiently reuses the subexpressions and costs a single multiplication rather than a complex exponential.

\begin{figure*}[t]
    \centering
    \includegraphics[trim={3.0cm 0 0 0},clip,width=\linewidth]{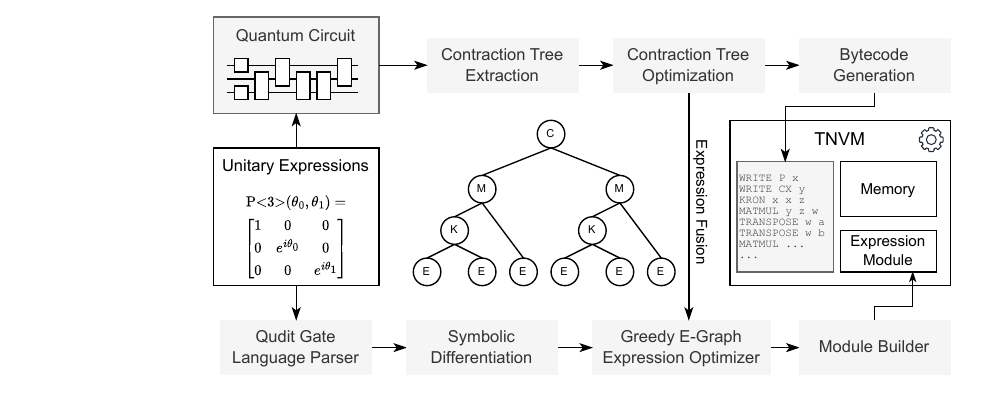}
    \caption{The OpenQudit compilation architecture, featuring two distinct pipelines. The ahead-of-time (AOT) pipeline (top) translates a quantum circuit into an optimized bytecode by solving the tensor network contraction ordering problem. The expression JIT pipeline (bottom), triggered during TNVM initialization, takes the QGL expressions referenced in the bytecode, performs symbolic differentiation and simplification, and compiles them into high-performance native functions for evaluation.}
    \label{fig:architecture}
\end{figure*}

\section{OpenQudit System Architecture}
\label{sec:openqudit}

The OpenQudit system architecture is divided into two primary phases: an ahead-of-time (AOT) compilation stage that produces a portable bytecode, and a runtime stage where a Tensor Network Virtual Machine (TNVM) executes this bytecode for rapid numerical evaluation. This end-to-end flow is illustrated in Figure~\ref{fig:architecture}, and a code example of this workflow is shown in Listing~\ref{lst:workflow_example}.

\begin{figure}[b!]
\begin{lstlisting}[
    style=livelyrust, 
    caption={
        The high-level user workflow in OpenQudit. The architecture separates the expensive, one-time setup from the fast, repeated evaluation loop: 
        (1) An ahead-of-time (AOT) pass compiles the PQC into an optimized bytecode. 
        (2) The TNVM is initialized once, allocating memory and JIT-compiling all expressions. 
        (3) The main optimization loop then consists of fast, repeated calls to the TNVM's `evaluate` method.
    },
    label={lst:workflow_example}
]
// (1) Ahead-of-time compilation (once per PQC)
let network = pqc.to_tensor_network();
let code = compile_network(network);

// (2) TNVM Initialization
let mut tnvm = TNVM::<c64, GRADIENT>::new(&code);

// (3) Fast evaluation loop
loop {
    let result = tnvm.evaluate(params);
    // ... update params using the result ...
}
\end{lstlisting}
\end{figure}

The process begins when a user defines a Parameterized Quantum Circuit (PQC) using the OpenQudit library and QGL expressions. The AOT compiler ingests this PQC, lowers it into a tensor network representation, and solves for an efficient contraction plan. This plan is then compiled into a specialized, optimized bytecode that fully describes the circuit's structure and the steps required to compute its unitary.

This bytecode is then passed to the TNVM to prepare for the numerical optimization loop. At initialization, the TNVM performs two key preparatory steps: it allocates all necessary memory buffers and eagerly JIT-compiles all unique QGL expressions referenced in the bytecode into high-performance native machine code. With these preparations complete, each subsequent call during the optimization loop simply involves the TNVM executing the highly-efficient bytecode to compute the PQC's unitary and its gradient.

\subsection{Ahead-of-Time Compilation to a Tensor Network Bytecode}

The first step in the AOT pipeline is to lower the PQC's structure into a \textit{tensor network} representation. In this model, each quantum gate becomes a tensor, and each input and output wire of the gate corresponds to an index of that tensor. For instance, a two-qubit gate, which connects two input and two output wires, is represented as a rank-4 tensor. The cardinality of each index corresponds to the dimension of the qudit on that wire---2 for a qubit, 3 for a qutrit, and so on. The wires connecting the gates in the circuit define the indices over which these tensors are contracted. The entire process of contracting the network results in a single tensor, which, after a final permutation and reshape, yields the unitary matrix of the full circuit.

\begin{table*}[t]
    \vspace{1mm}
    \centering
    \begin{tabular}{|l|l|p{9cm}|}
        \hline
        \textbf{Opcode} & \textbf{Operands} & \textbf{Description} \\
        \hline
        \texttt{WRITE} & \textit{expr\_id, buf\_out} & Evaluates the JIT-compiled QGL expression \textit{expr\_id}, writing the resulting matrix to buffer \textit{buf\_out}. \\
        \texttt{MATMUL} & \textit{buf\_in1, buf\_in2, buf\_out} & Performs matrix multiplication of matrices in \textit{buf\_in1} and \textit{buf\_in2}, storing the result in \textit{buf\_out}. \\
        \texttt{KRON} & \textit{buf\_in1, buf\_in2, buf\_out} & Performs a Kronecker product of matrices in \textit{buf\_in1} and \textit{buf\_in2}, storing the result in \textit{buf\_out}. \\
        \texttt{HADAMARD} & \textit{buf\_in1, buf\_in2, buf\_out} & Performs a Hadamard product (element-wise multiplication) of matrices in \textit{buf\_in1} and \textit{buf\_in2}, storing the result in \textit{buf\_out}. \\
        \texttt{TRANSPOSE} & \textit{buf\_in, shape, perm, buf\_out} & Fuses three operations: reshapes the input from \textit{buf\_in} into \textit{shape}, permutes the indices according to \textit{perm}, and reshapes back to a matrix in \textit{buf\_out}. \\
        \hline
    \end{tabular}
    \caption{The Tensor Network Virtual Machine (TNVM) bytecode instruction set. Operations act on abstract, labeled buffers.}
    \label{tab:bytecode}
\end{table*}

The computational cost of this contraction is critically dependent on the order in which the tensors are multiplied, known as the \textit{contraction path}. Contracting smaller intermediate tensors first can reduce the total floating-point operations by orders of magnitude. However, finding the optimal contraction path is an NP-hard problem~\cite{chi1997optimizing}. To balance compilation time with output quality, OpenQudit employs a hybrid strategy: an optimal solver~\cite{pfeifer2014faster} is used for small tensor networks where it is tractable, while a fast, greedy heuristic~\cite{gray2021hyper} is applied to networks with more than 7 tensors. This ensures that the AOT compilation itself does not become a significant performance bottleneck.

Once the optimal contraction path is determined, it is materialized as a binary \textit{contraction tree}. Each leaf node in this tree is an initial tensor representing a gate, and each internal node represents a pairwise contraction. During the construction of this tree, any required \texttt{trace} operations are handled immediately; they are applied symbolically to the QGL expressions at the leaf nodes to produce new, pre-traced expressions for the JIT stage. This design simplifies the contraction plan by eliminating the need for the bytecode to provide \texttt{trace} capabilities. After the initial tree is materialized, a separate optimization pass performs fusions to reduce the number of explicit runtime operations. For example, a \texttt{transpose} operation on a leaf node is fused by pushing it into the gate's symbolic QGL expression, allowing the JIT compiler to generate code for the already-transposed matrix directly.

After this optimization pass, the compiler analyzes the tree for parameter dependencies and serializes it into a two-section bytecode. Subtrees that do not depend on any PQC parameters are compiled into a \textit{constant} section, which the TNVM executes only once at initialization. The remaining parameter-dependent parts of the tree form the \textit{dynamic} section, which is executed on every call to the TNVM. The final bytecode is generated via a depth-first traversal of the tree, producing a sequence of instructions (e.g., \texttt{MATMUL}, \texttt{KRON}) that operate on abstract, labeled buffers. Each numerical contraction is scheduled using a high-performance transpose-transpose-GEMM-transpose (TTGT) strategy---a method well-suited for the types of problems targeted by OpenQudit on a CPU.

The final output of the AOT compiler is a \textit{bytecode} program consisting of a sequence of operations that act on these abstract buffers. For example, an instruction like \texttt{MATMUL 5 7 13} directs the TNVM to multiply the tensors currently in buffers 5 and 7, storing the result in buffer 13. This bytecode provides a portable, low-level representation of the entire contraction plan. The complete instruction set is detailed in Table~\ref{tab:bytecode}.

\subsection{The Tensor Network Virtual Machine (TNVM)}

The \textit{Tensor Network Virtual Machine (TNVM)} is a lightweight, high-performance runtime designed to execute the bytecode generated by the AOT compiler. A user instantiates a TNVM for a specific numerical task, configuring it with generic types for the desired precision (\texttt{f32} or \texttt{f64}) and the required level of differentiation (e.g., none, gradient, or Hessian).

This instantiation triggers a series of powerful, one-time preparatory steps to ensure the subsequent evaluation loop is maximally efficient. First, the TNVM allocates a single, contiguous memory region to house all intermediate tensors, eliminating dynamic allocation overhead during execution. Second, the TNVM eagerly JIT-compiles all unique QGL expressions from the \texttt{WRITE} instructions into native function pointers. Finally, it immediately computes any constant, non-parameterized sub-graphs identified in the bytecode, storing their results in read-only buffers.

\begin{figure*}[t]
    \centering
    \begin{subfigure}[b]{0.48\textwidth}
        \centering
        \includegraphics[width=\textwidth]{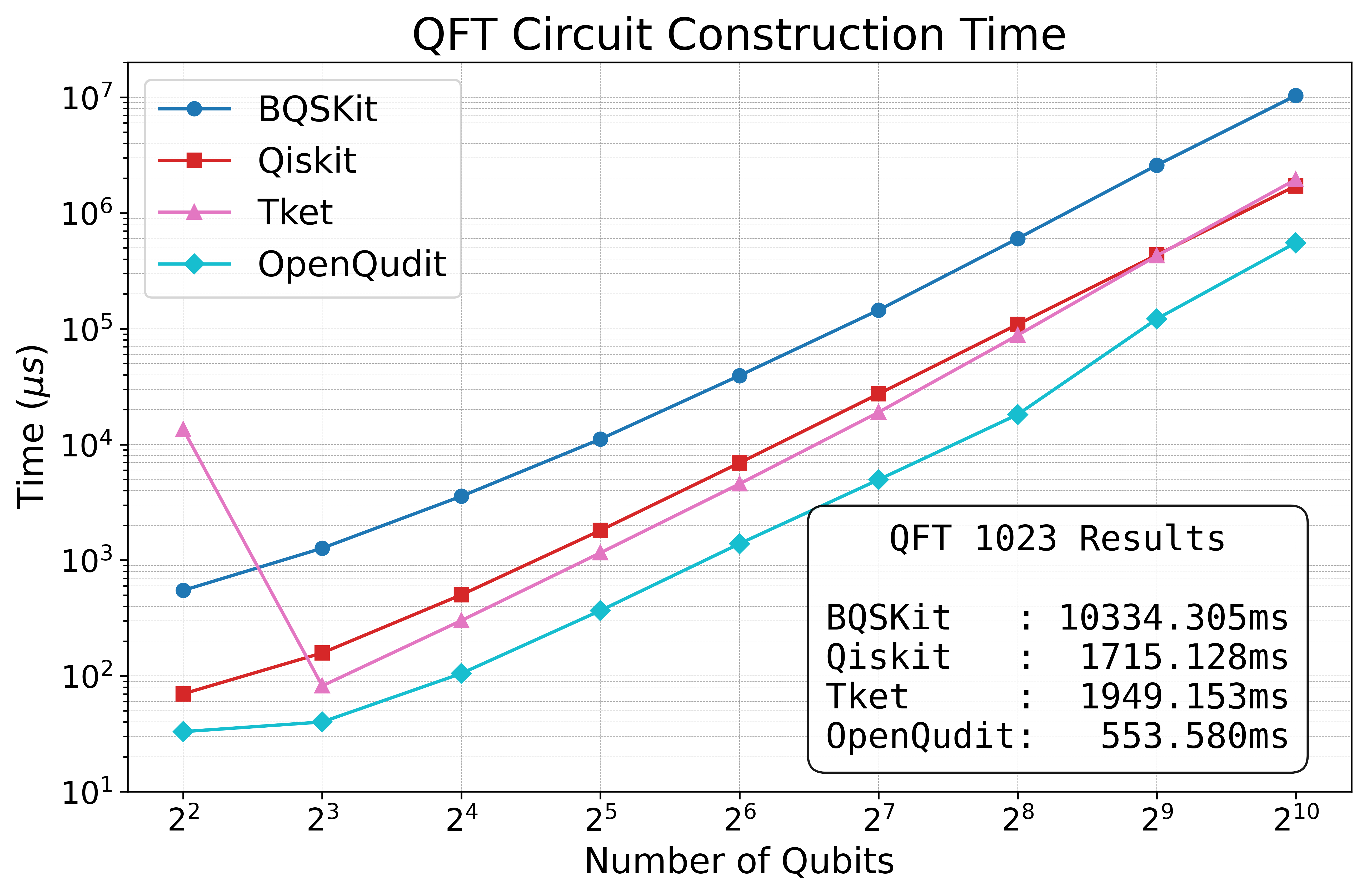}
        \label{fig:qft_construction}
    \end{subfigure}
    \hfill 
    \begin{subfigure}[b]{0.48\textwidth}
        \centering
        \includegraphics[width=\textwidth]{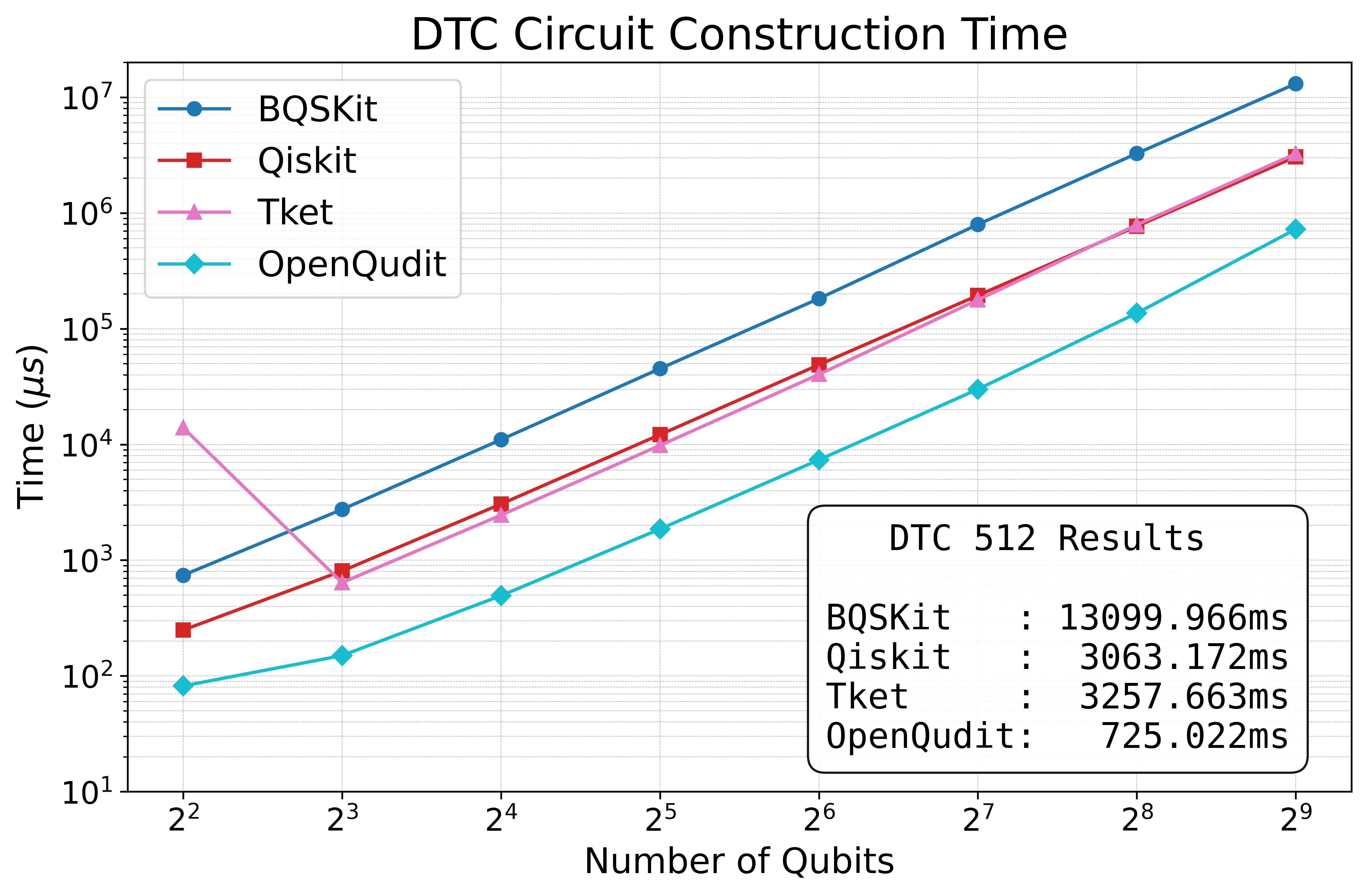}
        \label{fig:dtc_construction}
    \end{subfigure}
    
    \caption{
        Circuit construction time in OpenQudit compared to other frameworks for two scalable benchmarks, the Quantum Fourier Transform (QFT), and the DTC benchmark circuit from the Benchpress suite. Benchmarks were run on powers of two, except the final QFT circuit. QFT 1023 was benchmarked instead to avoid potential floating-point issues at 1024 qubits.
    }
    \label{fig:construction_performance}
\end{figure*}

A key challenge is that the JIT compilation of a single QGL expression with LLVM can take milliseconds, whereas a single numerical evaluation of the resulting circuit may only take microseconds. To amortize this cost, OpenQudit employs a caching mechanism. An \texttt{ExpressionCache} object is attached to each circuit and managed as shared state. Since the set of parameterized gates is typically fixed for a given quantum synthesis or compilation task, the \texttt{ExpressionCache} ensures that each unique QGL expression is JIT-compiled only once. The resulting LLVM context and compiled function pointers are cached, allowing subsequent TNVM initializations to retrieve the pre-compiled functions via a fast lookup. This upfront preparation ensures that the evaluation loop is free from allocation or compilation overhead.

To compute gradients beyond the QGL level, the TNVM is designed to support \textit{forward-mode automatic differentiation (AD)}. To facilitate this, the AOT compiler annotates each bytecode instruction with the set of symbolic parameters on which it depends. At initialization, the TNVM uses this information to specialize each instruction for the requested differentiation task. This allows the VM to correctly apply the rules of calculus, distinguishing between operations on independent partials (e.g., $\frac{\partial}{\partial x} [F(x)G(y)]$) and those with overlapping parameters that require the product rule (e.g., $\frac{\partial}{\partial x} [F(x)G(x)]$).

Once initialized, the evaluation of the PQC is straightforward and fast. The TNVM simply iterates through the linear list of specialized bytecode operations. Each instruction reads from its input buffers and writes to a designated output buffer, with the final result residing in a single, pre-determined buffer at the end of the sequence.

\section{Evaluation}
\label{sec:eval}

This evaluation answers two key questions. First, what is the overhead of a highly extensible, expression-based circuit representation? The results demonstrate that by using an expression caching mechanism, large-scale circuit construction in OpenQudit is significantly faster than in other state-of-the-art quantum compiler frameworks. Second, how much does OpenQudit accelerate the core numerical instantiation task? The experiments show that the OpenQudit compilation pipeline provides a significant speedup over BQSKit~\cite{bqskit}, the state-of-the-art numerical quantum compiler, on a variety of representative problems.

\begin{figure*}[t]
    \centering
    \includegraphics[width=\textwidth]{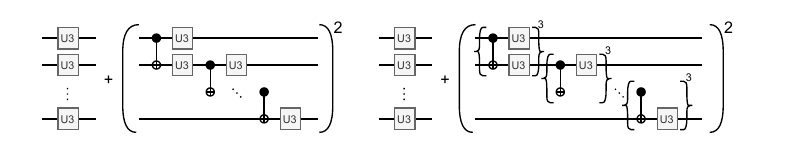}
    \caption{The set of PQC circuits used for the numerical instantiation benchmarks, varying in qudit count, radix, and depth. On the left, the shallow qubit circuit is illustrated, and on the right, the deep qubit circuit is illustrated. The qutrit version of the shallow circuit is similar in structure to the qubit one, except CSUM and Qutrit Phase gates are used instead of CNOT and U3 gates. These circuits are representative of candidate PQCs found during different stages of a numerical synthesis algorithm.}
    \label{fig:instantiation-circuits}
\end{figure*}

\begin{figure}[b!]
\begin{lstlisting}[
    style=livelyrust, 
    caption={
        A fully-encapsulated method to build a DTC circuit using the OpenQudit library. All gates used in the circuit are fully defined within the constructor. The constructor caches the gates, so that they can be appended to the circuit by reference.
    },
    label={lst:dtc_example}
]
fn build_dtc_circuit(n: usize) -> QuditCircuit {
  // Define gate using QGL's natural syntax
  let rx = UnitaryExpression::new("RX(theta) { 
    [[cos(theta/2), ~i*sin(theta/2)], 
     [~i*sin(theta/2), cos(theta/2)]]
  }");

  let rzz = UnitaryExpression::new("RZZ(theta) {
     [[e^(~i*theta/2), 0, 0, 0],
      [0, e^(i*theta/2), 0, 0],
      [0, 0, e^(i*theta/2), 0],
      [0, 0, 0, e^(~i*theta/2)]]
  }");

  let rz = UnitaryExpression::new("RZ(theta) {
    [[e^(~i*theta/2), 0],
     [0, e^(i*theta/2)]]
  }");

  // Initialize circuit and cache the expressions
  let mut circ = QuditCircuit::pure(vec![2; n]);
  let rx_ref = circ.cache_operation(rx);
  let rz_ref = circ.cache_operation(rz);
  let rzz_ref = circ.cache_operation(rzz);

  // Build the circuit
  for _ in 0..n {
    for i in 0..n {
      circ.append_ref_constant(
        rx_ref,         // Expression
        i,              // Qudits
        vec![0.95 * PI] // Parameters
      );
    }
    // ... other loops omitted ...
  }
  circ
}
\end{lstlisting}
\end{figure}

\subsection{Experimental Setup}
All experiments were conducted on a single core of an AMD 5800X processor at 4.85\,GHz in a machine with 64\,GB of memory. The OpenQudit framework, implemented in approximately 20,000 lines of Rust, is publicly available at \url{https://github.com/openqudit/openqudit}, and all benchmarks and scripts used in the evaluation are publicly available on Zenodo~\cite{openqudit-artifact}. OpenQudit uses LLVM version 18 for expression JIT compilation. Matrix multiplication and tensor transpose operations were performed by nano-gemm~\cite{nanogemm} and custom routines, respectively, while all other linear algebra used the faer library v0.22.6~\cite{faer}.

OpenQudit is compared against leading Python-based quantum compilers: BQSKit 1.2.1, Qiskit 2.1.2~\cite{Qiskit}, and Tket 2.9.3~\cite{tket}, using Python 3.13.7.

\begin{figure*}[t]
    \centering
    \includegraphics[width=\textwidth]{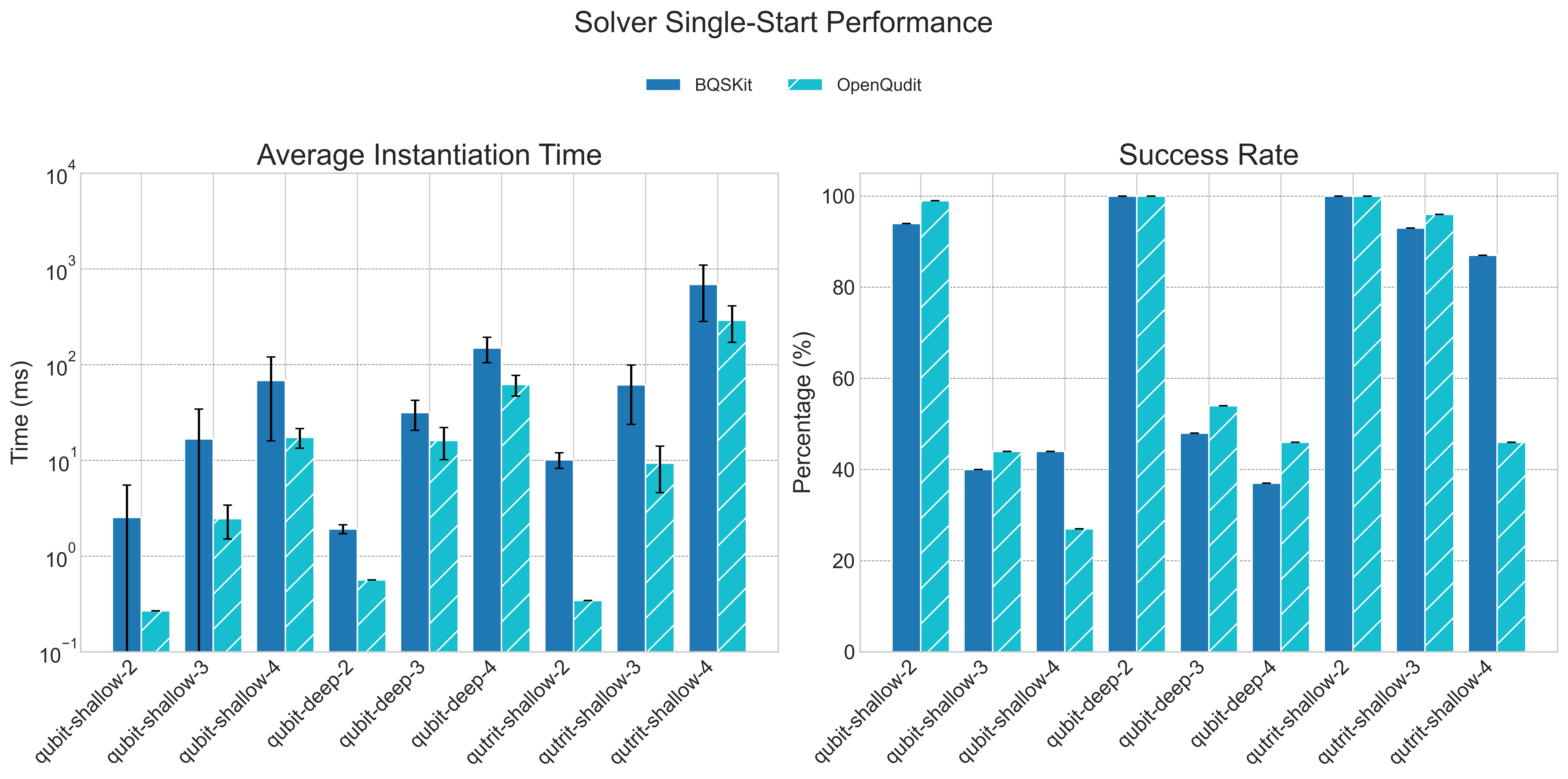}
    \caption{(Left) Instantiation time for a single optimization run. (Right) The corresponding success rate. OpenQudit provides a speedup of \textbf{6.8x} on 3-qubit shallow circuits and \textbf{6.6x} on 3-qutrit circuits.}
    \label{fig:singlestart}
\end{figure*}

\begin{figure*}[h!]
    \centering
    \includegraphics[width=\textwidth]{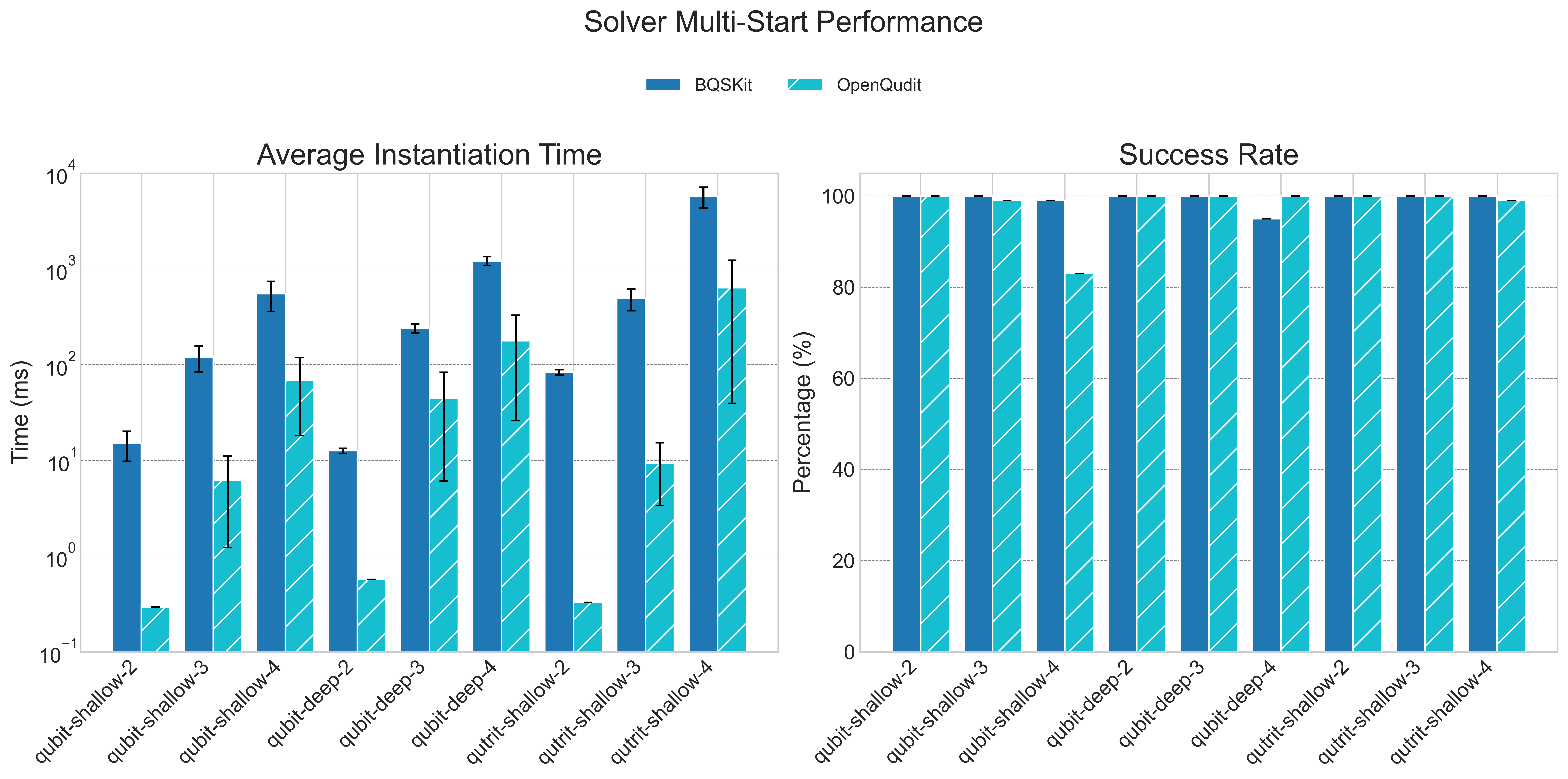}
    \caption{(Left) Total instantiation time for 8 optimization runs. (Right) The success rate approaches 100\% for most benchmarks. The benefits of OpenQudit's AOT compilation are amortized, yielding a \textbf{19.6x} speedup on the common 3-qubit shallow circuit case.}
    \label{fig:multistart}
\end{figure*}

\subsection{Circuit Construction Performance}
A fast and flexible API for programmatic circuit construction is essential for any compiler framework. To measure this, the time required to build two scalable quantum circuits of varying sizes was benchmarked: the \textit{Quantum Fourier Transform (QFT)}~\cite{qft_orign}, a structured and ubiquitous component, and a \textit{Discrete Time Crystal (DTC)} hamiltonian simulation circuit~\cite{zhang2023characterizing}, a benchmark from the Benchpress~\cite{nation2024benchmarking} benchmark suite containing random parameters.

As shown in Figure~\ref{fig:construction_performance}, OpenQudit significantly outperforms all baselines in large-scale circuit construction. This performance is due to its \textit{expression caching} mechanism, demonstrated in Listing~\ref{lst:dtc_example}. By defining a gate's semantics with QGL only once and subsequently adding it to the circuit via a lightweight integer reference, OpenQudit avoids repeated and costly, but necessary, safety and equality checks. For example, the 1023-qubit QFT circuit is constructed in under a second, a task that takes all other frameworks significantly longer. OpenQudit builds the 512-qubit DTC circuit \textbf{18.1x} faster than BQSKit and over \textbf{4x} faster than Qiskit and Tket.

\subsection{Numerical Instantiation Performance}
The core instantiation performance of OpenQudit is evaluated against BQSKit, the only one of the tested frameworks with a built-in numerical compilation engine. BQSKit's engine is highly optimized and uses Google CERES's~\cite{Agarwal_Ceres_Solver_2022} Levenberg-Marquardt~\cite{levenberg1944method, marquardt1963algorithm} (LM) algorithm. A naive LM optimizer was implemented within OpenQudit that leverages the TNVM for all unitary and gradient calculations. For a fair comparison, the reported timings for OpenQudit include the one-time AOT compilation cost, a step BQSKit does not need to perform.

A set of PQC circuits representative of those found in synthesis algorithms like QSearch~\cite{qsearch} was benchmarked, varying the number of qudits, circuit depth, and radix (Figure~\ref{fig:instantiation-circuits}). Due to the complex quantum optimization landscape, a common technique to improve the success rate is to run the optimizer multiple times from different random starting parameters (\textit{multi-start}). Both single-start performance and the more realistic multi-start scenario are evaluated, using 8 starts to match the default for BQSKit's most common optimization level (\texttt{-O3}).

The results are shown in Figures~\ref{fig:singlestart} and~\ref{fig:multistart}. In the single-start case, OpenQudit provides a consistent speedup across all benchmarks, including a \textbf{6.8x} speedup for shallow 3-qubit circuits and a \textbf{6.6x} speedup for 3-qutrit circuits, demonstrating the efficiency of the system architecture. 

The performance advantage is even more pronounced in the multi-start scenario. The cost of OpenQudit's AOT compilation is amortized over the multiple runs, in addition to short-circuiting by terminating early once a valid solution is found. This results in a \textbf{19.6x} speedup for the 3-qubit shallow case, a task that is a frequent subroutine in larger synthesis algorithms. In all cases, OpenQudit provides a significant performance improvement over the state-of-the-art. While the success rate is competitive across the board, OpenQudit's naive implementation of the LM algorithm shows room for improvement compared to the well-established CERES library. Furthermore, the TNVM maintains a low memory footprint; for example, the 3-qubit shallow circuit benchmarks required only 211KB of memory for double-precision evaluation.

\section{Discussion}
\label{sec:disc}

\subsection{Performance and the Role of the Optimizer}
The evaluation in this paper utilized a naive implementation of the Levenberg-Marquardt (LM) optimizer. This was a deliberate choice intended to isolate and measure the performance of the underlying unitary and gradient calculations executed by the TNVM, which is the core contribution of this work. The results show that even with a simple optimizer, the TNVM provides a significant speedup over the highly tuned engine in BQSKit. The reported speedups, achieved even with a naive Levenberg-Marquardt (LM) implementation, indicate that significant further acceleration is possible by developing a more sophisticated optimizer built directly on the TNVM.

\subsection{The AOT Compilation Trade-Off}
The OpenQudit architecture makes a clear trade-off: it invests in an upfront, ahead-of-time (AOT) compilation cost to enable an extremely fast runtime. While this AOT step, which includes tensor network lowering and contraction pathfinding, is non-trivial, its cost is paid only once. As demonstrated by the multi-start instantiation results (Figure~\ref{fig:multistart}), this cost is quickly amortized in typical numerical synthesis workloads, which require many evaluations. This AOT/runtime hybrid model is thus well-suited to the high-throughput demands of its target domain.

\subsection{Discussion on Floating-Point Precision}
All experiments in this work were conducted using double-precision floating-point numbers (\texttt{f64}) to ensure a fair and stable comparison against baselines. While the OpenQudit TNVM is generic and can be instantiated for single-precision (\texttt{f32}), the naive Levenberg-Marquardt (LM) optimizer implemented for the evaluation is not sophisticated enough to converge reliably with lower precision without error.

However, this presents a significant opportunity for future performance gains. Numerical optimization using \texttt{f32} is typically much faster. TNVM gradient evaluations of the 3-qubit shallow circuit show an $1.27\times$ speedup when run with f32 vs f64 precision (25.59 vs 32.579 microseconds). A key architectural advantage of OpenQudit is its flexibility in this regard; the precision is a generic parameter of the TNVM. This is in contrast to many existing frameworks, like BQSKit, which do not expose this level of control. Future work could explore the integration of a more robust, precision-aware optimization algorithm.

\subsection{Future Directions}
The symbolic design of QGL and the OpenQudit IR provides a clear path for several future extensions. The most significant is the extension of QGL to support dynamic circuits by incorporating definitions for non-unitary operations such as \texttt{state}, \texttt{isometry}, and \texttt{Kraus} operators. This would enable the framework to handle state preparations, mid-circuit measurements, and noise models. Furthermore, the TNVM could be extended to target different hardware backends. While the current implementation is optimized for CPUs, the bytecode is platform-agnostic, and a future version could generate specialized code for GPUs, potentially leveraging high-performance libraries like cuTensor~\cite{cuquantum}.

\section{Related Work}
\label{sec:related}

\subsection{Numerical Quantum Frameworks}

Several powerful frameworks exist for the numerical evaluation or optimization of quantum programs, including \textit{PennyLane}~\cite{bergholm2018pennylane}, \textit{TorchQuantum}~\cite{wang2022torchquantum}, and \textit{JAX-Quantum}~\cite{jha2024jaxquantum}. These tools are built on top of popular frameworks like JAX~\cite{jax2018github} or PyTorch~\cite{imambi2021pytorch} and are often designed for applications such as quantum machine learning, variational quantum algorithms, or state-vector simulation. These workloads are categorically different from OpenQudit's and are ideal use-cases for these existing frameworks, since they focus on one or a relatively small number of memory-bound, large-scale simulations.

Conversely, these general-purpose frameworks are not well-suited for OpenQudit's domain due to their higher in-the-loop cost. For example, a JIT-compiled U3 gate evaluation takes $\sim$6$\mu$s with JAX, while it takes $<100$ns with OpenQudit. Numerical quantum compilation is a high-throughput task, requiring extremely fast classical optimization subroutines, for which efficient analytic gradients are essential at the gate-level. Since typical compilations require millions to billions of optimization calls, the one-time compilation cost of OpenQudit is quickly amortized. This is the specific domain that OpenQudit is designed to accelerate.


\subsection{Quantum Programming Languages}
Most high-level quantum languages, such as \textit{Q\#}~\cite{qsharp} or \textit{OpenQASM}~\cite{openqasm}, provide an API for constructing circuits from a predefined, hard-coded set of gate primitives. QGL appears to be the first to allow users to define the semantics of a new, low-level instruction directly as a symbolic mathematical expression. While other mathematical libraries for representing quantum operators exist, they typically require users to construct these operators by composing a sequence of code-based function calls. QGL's design, in contrast, provides a direct, declarative syntax that mirrors on-paper mathematical notation. It is important to note, however, that QGL is not a complete, standalone quantum language for writing full algorithms. It can be embedded into other languages to provide semantic defining features.

\subsection{Tensor Network Compilation}
The problem of finding an optimal tensor network contraction path is critical for performance. State-of-the-art libraries like Cotengra~\cite{gray2021hyper} provide highly optimized solvers for this specific NP-hard problem. Furthermore, specialized libraries like NVIDIA's cuTensor~\cite{cuquantum} provide highly optimized GPU kernels for tensor network contraction. While OpenQudit incorporates similar pathfinding heuristics, its primary contribution is not the contraction kernel itself. Rather, OpenQudit provides an end-to-end compilation system---including an AOT compiler and a specialized virtual machine---purpose-built to accelerate the entire PQC instantiation workload.

\subsection{Quantum Circuit Simulators}
High-performance quantum circuit simulators, such as QuEST~\cite{jones2019quest} and NVIDIA's cuQuantum~\cite{cuquantum}, represent another related area of research. The primary objective of these frameworks is to classically simulate the evolution of a quantum state under the action of a large quantum circuit, often leveraging distributed high-performance computing (HPC) systems to manage the exponential complexity. This focus is distinct from that of OpenQudit. While simulators are optimized for the scale and fidelity of a single, large-scale execution---typically without regard for gradients---OpenQudit is designed for the high-throughput, repeated unitary evaluation of many small circuits and their gradients.

\section{Conclusion}
\label{sec:conc}

Numerical quantum compilers, while powerful, are fundamentally limited by the performance of their classical optimization subroutines and a design that makes extending them with new gate sets a difficult, error-prone task. This paper introduced OpenQudit, a compilation framework that addresses these challenges by treating quantum operations as symbolic expressions. By combining the Qudit Gate Language (QGL) for defining gate semantics with a compilation pipeline that uses e-graphs for simplification and a tensor network virtual machine for execution, OpenQudit separates the mathematical definition of an operation from its high-performance implementation. The evaluation demonstrated that this symbolic compiled approach is highly effective, accelerating the core numerical instantiation task on common, high-throughput synthesis workloads. Ultimately, OpenQudit provides a path towards more performant and accessible numerical compilers, enabling researchers to more rapidly explore and deploy novel quantum operations.

\section*{Acknowledgment}
This work was supported by the U.S. Department of Energy, Office of Science, Office of Advanced Scientific Computing Research under Contract No. DE-AC05-00OR22725 through the Accelerated Research in Quantum Computing Program MACH-Q project.

\bibliographystyle{IEEEtran}
\bibliography{quantum}

\end{document}